\documentclass[aps,twocolumn,showpacs,groupedaddress]{revtex4}
\usepackage{graphicx}

\begin{document}

\title{
Effect of Decoherence on the Dynamics of Bose-Einstein Condensates
in a Double-well Potential}
\author{W. Wang$^1$, L. B. Fu$^2$, and  X. X. Yi$^1$}

\email{yixx@dlut.edu.cn}

\affiliation{$^1$Department of Physics, Dalian University of
Technology, Dalian 116024, China\\
$^2$Institute of Applied Physics and Computational Mathematics,
Beijing 100088, China\\}
\date{\today}

\begin{abstract}
We study the dynamics of a Bose-Einstein condensate in a double-well
potential in the mean-field approximation. Decoherence effects are
considered by analyzing the couplings of the condensate to
environments. Two kinds of coupling are taken into account. With the
first kind of coupling dominated, the decoherence can enhance the
self-trapping by increasing the damping of the oscillations in the
dynamics, while the decoherence from the second kind of
condensate-environment coupling leads to   spoiling of the quantum
tunneling and self-trapping.
\end{abstract}

 \maketitle

Bose-Einstein condensates(BECs) in a double-well potential exhibit
many fascinating phenomena that are absent in thermal atomic
ensembles, for example, quantum tunneling and
self-trapping\cite{milburn97, smerzi97,
raghavan99,kohler02,salasnich00, wang06}. Quantum tunneling through
a barrier is a paradigm of quantum mechanics and usually takes place
on a nanoscopic scale, such as in two supperconductors separated by
a thin insulator\cite{likharev79} and two reservoirs  of superfluid
helium connected by nanoscopic apertures\cite{pereverzev97,
sukhatme01}. Recently, tunneling on a macroscopic scale ($\mu m$) in
two weakly linked Bose-Einstein condensates in a double-well
potential has been observed\cite{albiez05}. Similar to tunneling
oscillations in superconducting and superfluid Josephson junctions,
Josephson oscillations  are observed when the initial population
difference is chosen to be below the critical value. When the
initial population difference exceeds a critical value, an
interesting feature of the coherent quantum tunneling between the
two BECs is observed, i.e., tunneling oscillations are suppressed
due to the nonlinear condensate self-interactions. This phenomenon
is known as macroscopic quantum self-trapping.

The interactions between the condensate and noncondensate atoms
lead to decoherence. Describing decoherence by fully including the
quantum effects requires sophisticated theoretical studies that
include the effect of noncondensate atoms. Treating the
noncondensate atoms as a Markovian reservoir, master equations
that govern the dynamics of the condensate atoms might be derived
\cite{dalvit00,louis01,micheli03}. In fact, in the experiments on
BECs, trapped atoms are evaporatively cooled and they continuously
exchange particles with the noncondensate atoms. Thus standard
procedure in quantum optics for open systems would naturely lead
to master equations for treating atomic BECs. This Markovian
treatment for the BECs also can be understood as the presence of
lasers for trapping/detection of atoms, which will polarize the
atoms and thus couple them to the vacuum modes of the
electromagnetic field\cite{huang06}. On the other hand, due to the
unavoidable interaction of the BECs with its environment, the
decoherence is always there in BECs, hence the characterization of
decoherence in this  system become interesting.  Because different
decoherence may have different effects on the dynamics of the
BECs, the character of decoherence in the BECs may be read out
from the dynamics of the BECs. Indeed, as we shall show, different
BEC-environment coupling leads to different final population
imbalance of the BECs in the double-well potential. This may be
used to characterize the decoherence in the double-well systems.

In this paper, we study the effect of decoherence on the dynamics of
BEC in a double-well potential by studying the  evolution of the
master equation for the  BEC  within a mean-field framework,  where
the number of atoms in the condensates is supposed to be infinity
and the quantum fluctuation is negligible. To derive the master
equation, we need modeling the environment and BEC-environment
coupling. However, this is not a easy task that we do not address at
present. Instead, we write the master equation by analyzing the
effects of environmently induced decoherence. When analyzing
decoherence effects on the dynamics of BECs in a double-well
potential, we are interested in answering two basic
questions:(1)What effects are made by the decoherence on
self-trapping in the BECs? And (2) how does the decoherence affect
the quantum tunneling in BECs in a double-well potential?

Consider BECs in a double-well potential, the wave function of BECs
can be expressed as the supperposition of individual wave functions
in each well,
\begin{equation}
|\phi\rangle=a_R|R\rangle+a_L|L\rangle,
\end{equation}
where $|R\rangle$ and $|L\rangle$ denote the wavefunction of the
right and left well, respectively. The coefficients $a_R$ and $a_L$
of the expansion satisfy the two-mode Gross-Pitaevskii equation
(GPE)\cite{milburn97, smerzi97}(setting $\hbar=1$),
\begin{equation}
i\frac{\partial}{\partial t}\left ( \matrix{a_R \cr a_L}\right )=
H\left( \matrix{a_R \cr a_L}\right ).\label{nls1}
\end{equation}
The Hamiltonian is given by

\begin{eqnarray}
H= \left( \matrix{ \frac {\gamma}{2}+\frac{c}{2}(|a_R|^2-|a_L|^2)
& \frac V 2
  \cr
  \frac{V}{2} & -\frac {\gamma}{2}-\frac{c}{2}(|a_R|^2-|a_L|^2)
   \cr } \right),\nonumber\\
   \label{nlh1}
\end{eqnarray}
where $\gamma$ is the energy bias between the two wells, $c$
stands for the nonlinear parameter describing the condensate
self-interaction, and $V$ depending on the height of the barrier,
is the coupling constant between the two condensates. In this
paper, we shall focus on $\gamma=0$, i.e., the case of BECs in a
symmetric double-well potential. This situation is interesting
because the amplitude distributions of all eigenstates are
symmetric, leading to Josephson oscillations in the absence of
decoherence. With the Markov approximation, the master  equation
that results from the condensate-environment coupling takes the
form,
\begin{eqnarray}
i\frac{\partial}{\partial t}\rho &=&[H_{\rho},\rho]+{\cal
L}(\rho),\nonumber\\
{\cal L}(\rho)&=&i\frac{\Gamma}{2} (2A\rho A^{\dagger}-\rho
A^{\dagger}A-A^{\dagger}A\rho),\label{me}
\end{eqnarray}
where $\Gamma$ denotes the decoherence rate of the condensates,
and $A$ stands for an operator of the condensates. This master
equation can be derived by assuming that the
condensates-environment couplings take the form $H_I\sim \sum_j
g_j (Ab_j^{\dagger}+ h.c.),$ where $g_j$ denotes the constant of
interaction between the condensates and the environmental mode
$b_j$.  The condensate operator $A$ in general is expressed as a
linear superposition of three pauli operators, i.e.,
$A=\lambda_x\sigma_x+\lambda_y\sigma_y+\lambda_z\sigma_z$ with
notations $\sigma_z=|R\rangle\langle R|-|L\rangle\langle L|$,
$\sigma_x=|R\rangle\langle L|+|L\rangle\langle R|,$ and it is
similar for $\sigma_y$. The values of $\lambda_{\alpha}
(\alpha=x,y,z)$ depends on the source of decoherence and its
couplings to  the environment. For example,
$\lambda_y=\lambda_x=0$ is for the environment that dephasingly
couples to the condensates, while $\lambda_z=0$ is for the
environment leading the  BECs into dissipation. $H_{\rho}$ takes
the same form as in Eq.(\ref{nlh1}), except a change
$|a_x|^2\rightarrow \rho_{xx}=\langle x|\rho|x\rangle, x=R,L.$

To start with, we consider the case of
$A=\sigma_+=\sigma_x+i\sigma_y.$ This situation happens in the case
where the double-well potential is formed by using a Raman scheme to
couple two hyperfine states in a spinor BEC. The condensate in the
upper hyperfine states decays into the lower one, reminiscent of
atomic spontaneous emission.
\begin{figure}
\includegraphics*[width=0.8\columnwidth,
height=0.6\columnwidth]{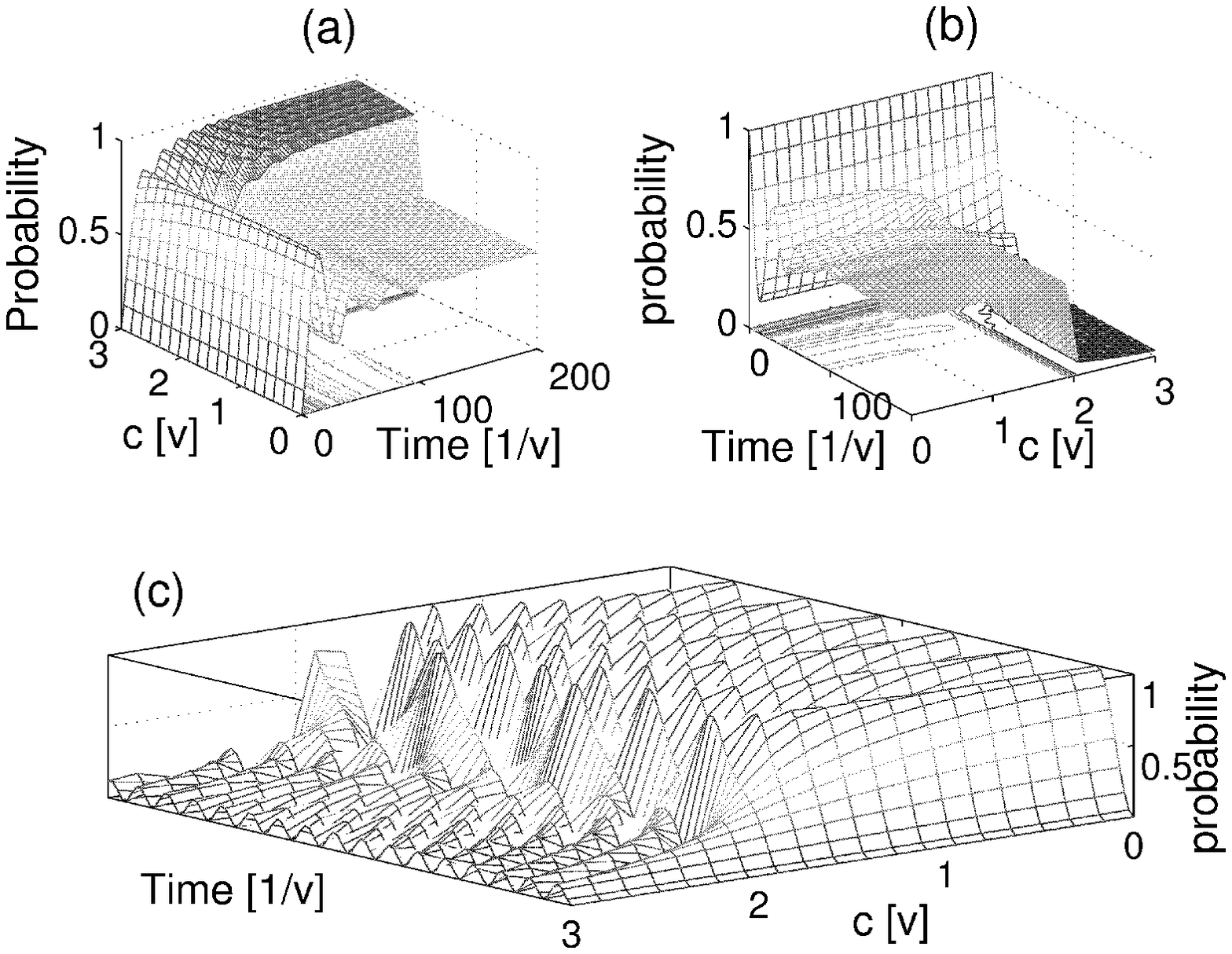} \caption{(color online)
Populations of the condensates in the Left well $|L\rangle$ [(a) and
(c)] and Right well $|R\rangle$ [(b)]. The condensates were
initially prepared in the Right well, the decoherence rate was
chosen $\Gamma=0.1 V$ in (a) and (b), while  $\Gamma=0$ in (c). A
jump-like change at $c=2V$ in 1-(a) and 1-(b) clearly appears due to
the decoherence effect. The nonlinear coupling constant was plotted
in units of $V$, while the time in units of $1/V$ in all figures in
this paper. } \label{fig1}
\end{figure}
The dynamics of the master equation is studied by numerical
simulations,  the results are presented in Fig. \ref{fig1} and
Fig. \ref{fig2}. In Fig. 1-(a) and 1-(b), we have plotted the
population of condensates in the Left well 1-(a) and Right will
1-(b). The initial state is all the condensate atoms in the Right
well, and the decoherence rate has been set to be $\Gamma=0.1V.$
In contrast, the dynamics of the condensate in the Left well
without decoherence ($\Gamma=0$) is presented in Fig.
\ref{fig1}-(c). Clearly, the decoherence increase the damping of
the oscillations. When the nonlinearity characterized by $c$ is
small compared to the tunneling $V$, the oscillations of the
population are suppressed, and the condensate finally remains in
the two wells with equal probability.
\begin{figure}
\includegraphics*[width=0.8\columnwidth,
height=1\columnwidth]{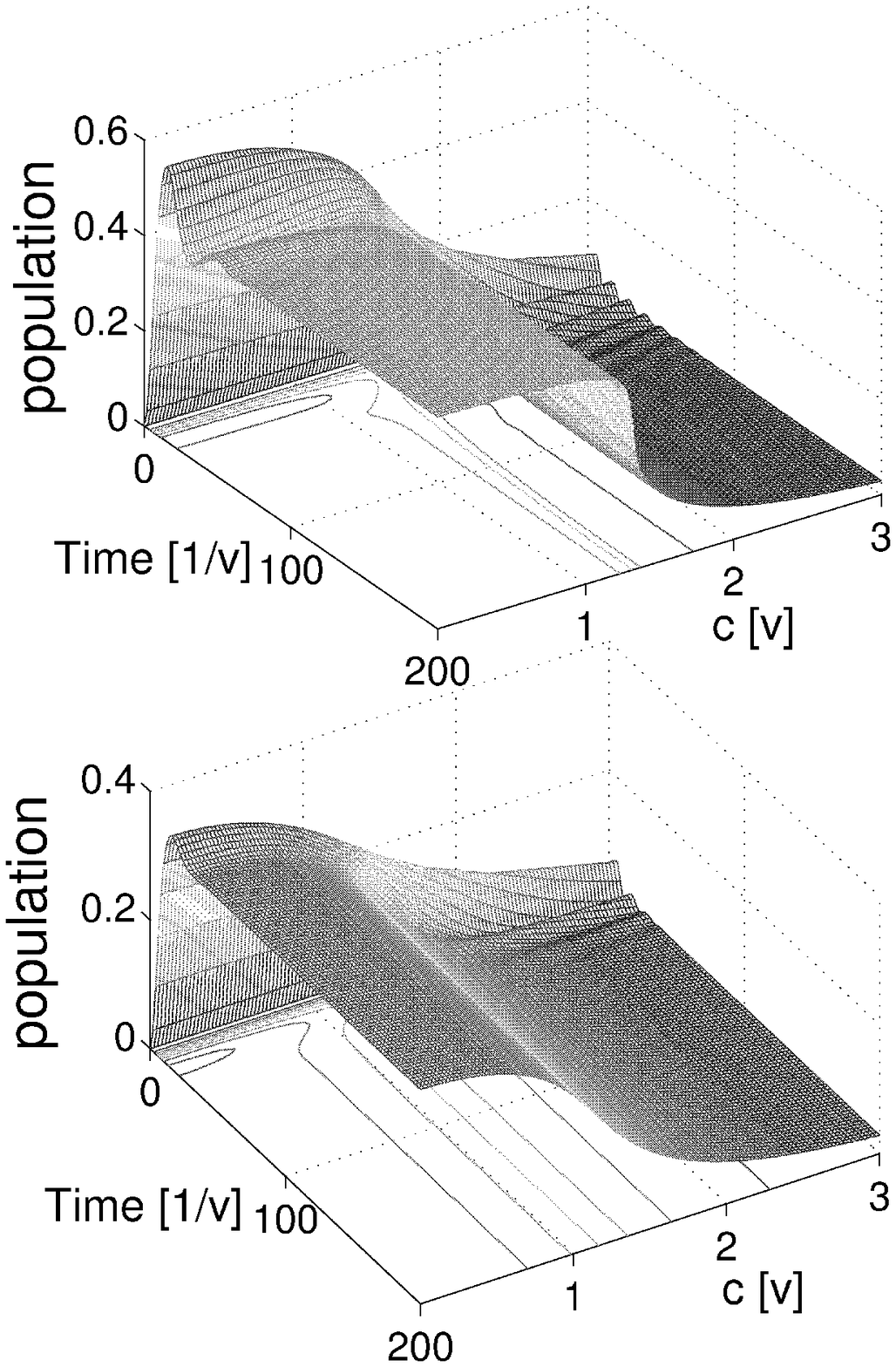} \caption{The same as
Fig.\ref{fig1}-(b) but with larger $\Gamma$, $\Gamma=0.3V$ for the
upper panel, while $\Gamma=0.5V$ for the lower panel. The jump-like
change in the population disappears with $\Gamma$ increasing. }
\label{fig2}
\end{figure}
If the nonlinearity is large with respect to the tunneling $V$ and
the population imbalance exceeds a critical value, the condensate
would be locked in one of the wells, depending on the initial
population. We would like to notice that the population change
drastically in the vicinity of the critical value $c=2V$, this is
due to the suppression of population oscillations by the
decoherence. With decoherence increasing, the jump-like change
near the critical value in the population becomes unclear, as
Fig.\ref{fig2} shows. That means the decoherence may determine the
final population imbalance in the two wells.
\begin{figure}
\includegraphics*[width=0.8\columnwidth,
height=0.6\columnwidth]{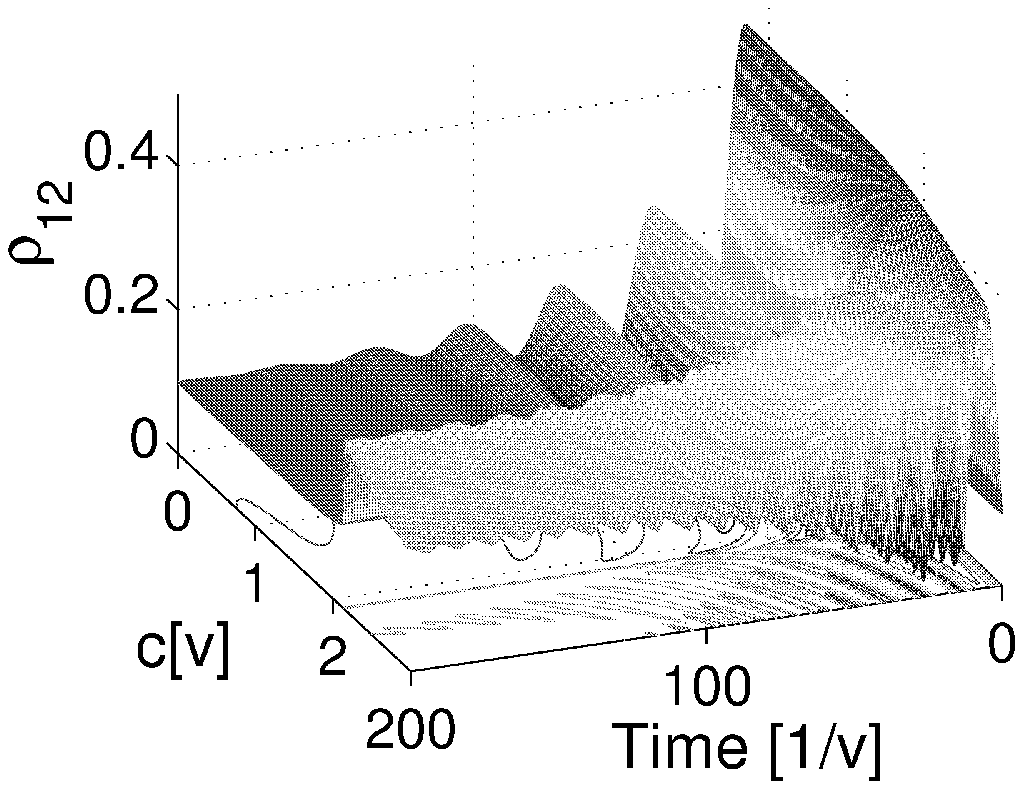} \caption{Norm of the
off-diagonal element of the density matrix $|\rho_{12}|$ as a
function of time and $c$, which is usually used to characterize the
decoherence.  The parameter chosen is $\Gamma=0.1V$, and the
condensates were initially in the Right well.} \label{fig3}
\end{figure}
On the other hand, the nonlinear interaction together with the
initial population and the relative phase can affect the
decoherence effect, which may be characterized by $|\rho_{12}|$,
i.e., the norm of the off-diagonal element of the density matrix.
This effect was shown in figure \ref{fig3}. We would like to note
that the jump-like change in Fig.\ref{fig1} might appear at
different $c$, depending on the fixed points around that the
population imbalance and the relative phase oscillate. For
example, our simulations show that the jump-like change could
appear at $c=V$ with initial relative phase $\theta=\pi$, and
non-zero population imbalance\cite{wang06}.

\begin{figure}
\includegraphics*[width=0.8\columnwidth,
height=0.6\columnwidth]{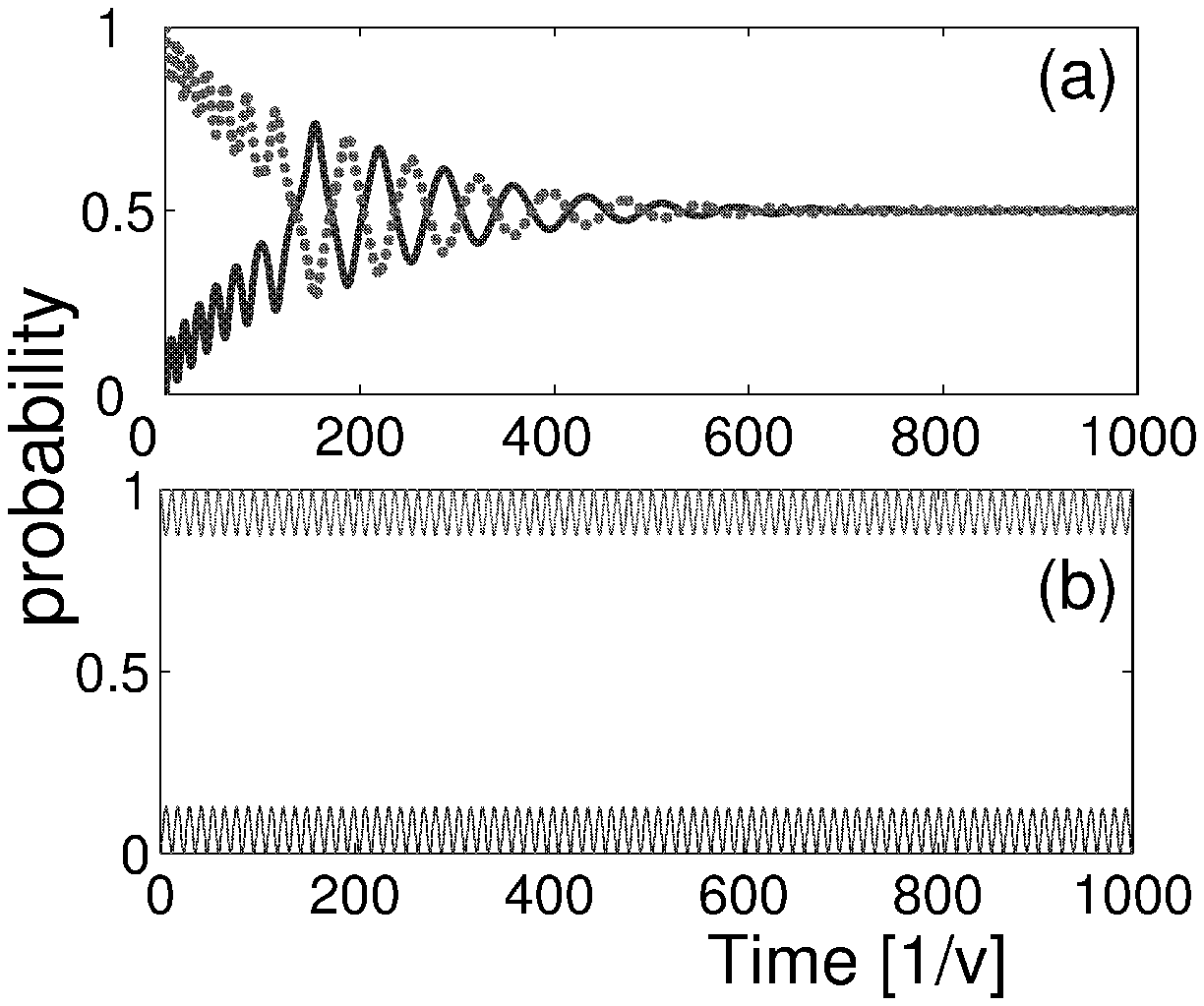} \caption{ (Color online)
Population of the condensates changes with time. The parameters
chosen are, (a) $c=3V$,  $\Gamma=0.01V$, and (b) $c=3V$,$\Gamma=0.$
Self-trapping occurs in the absence of decoherence, as figure 4-(b)
shows.
 The decoherence first drives the BECs from the self-trapping regime to the quantum
 tunneling regime, then it destroys the quantum tunneling. } \label{fig4}
\end{figure}

\begin{figure}
\includegraphics*[width=0.8\columnwidth,
height=0.6\columnwidth]{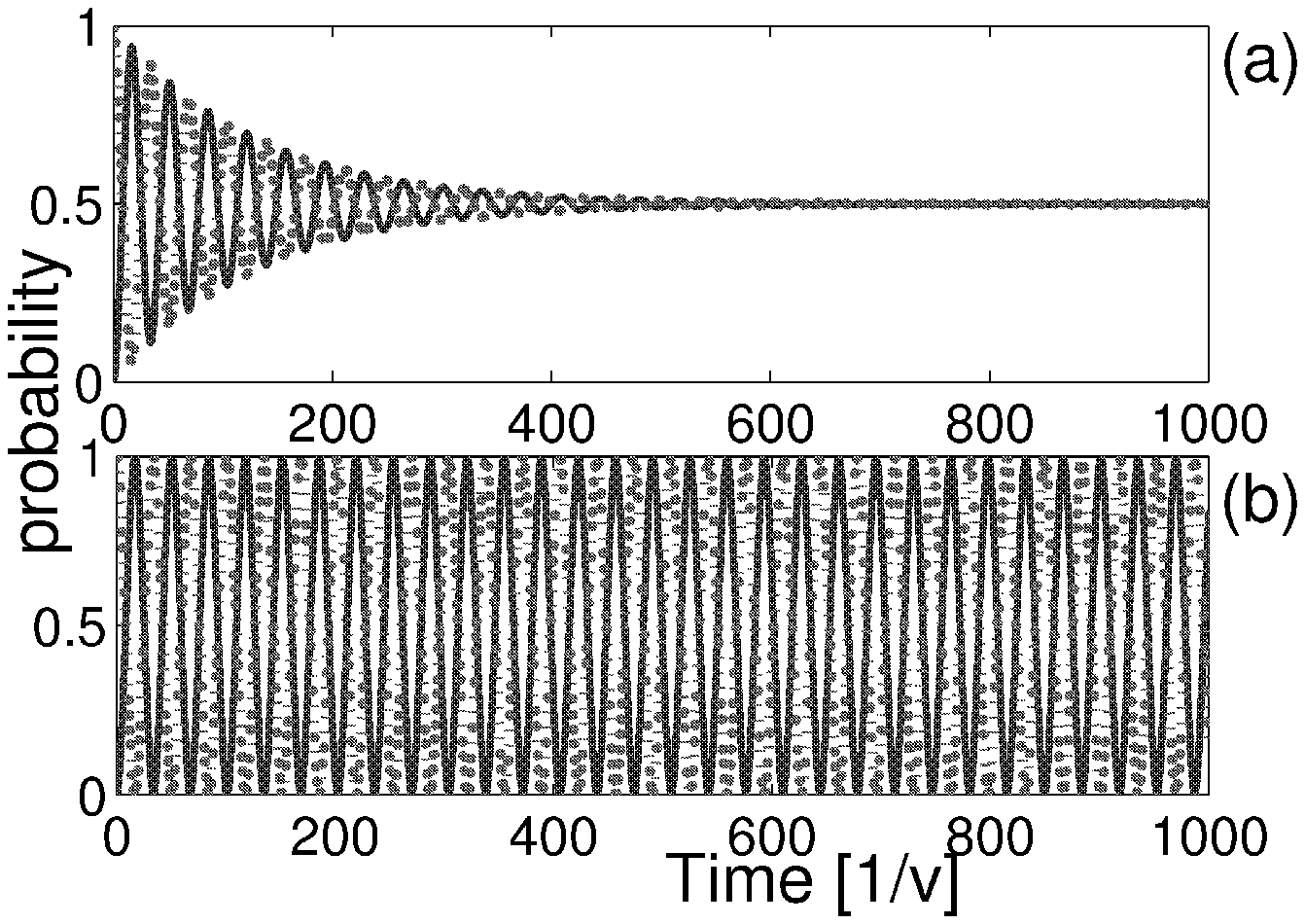} \caption{The same as
Fig.\ref{fig4}, but with $c=V.$ Solid line is plotted for the
population of BEC in the Right well, while dotted line for the
Left.} \label{fig5}
\end{figure}

\begin{figure}
\includegraphics*[width=0.8\columnwidth,
height=0.6\columnwidth]{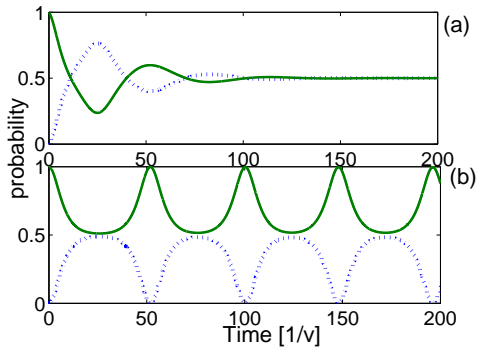} \caption{This figure shows
the population of the condensate at the critical value $c=2V.$ (a)
is plotted for $\Gamma=0.01V$, while (b) is for $\Gamma=0.$ It
confirms that the decoherence first leads the BEC from the
self-trapping regime to the quantum tunneling regime, then spoils
the quantum tunneling. } \label{fig6}
\end{figure}

Next, we take $A=\sigma_x$, corresponding to BECs in a spatial
double-well potential. The tunneling is driven by an environment
(or by fluctuational fields), leading to the decay in the quantum
tunneling. An alternative BEC system can be formed by using a
Raman scheme to couple two degenerate hyperfine states in a spinor
BEC. The driving fields may fluctuate, resulting in decoherence in
the quantum tunneling. We have performed extensive numerical
simulations for the master equation Eq.(\ref{me}). Selected
results, divided into three regimes by the nonlinearity, are
presented in Fig. \ref{fig4}, \ref{fig5}, and \ref{fig6}.
Fig.\ref{fig4} shows the dynamics of the condensate in the
self-trapping regime. The decoherence clearly increases the
amplitude of oscillations in the population  first, then increases
the damping of the amplitude of oscillations, meanwhile it
averagely decreases the population imbalance, and finally spoils
the self-trapping. In the self-trapping regime, the frequency of
the oscillation depends on the nonlinear parameter $c$, the
initial population imbalance and relative phase, as well as the
coupling constant $V$ between the BECs. This can be found by
comparing Fig.\ref{fig4}, \ref{fig5} and \ref{fig6}. With $c$ and
$V$ fixed, the decoherence changes the population imbalance, this
results in the frequency change as shown in Fig.\ref{fig4}-(a).
In the quantum tunneling regime(Fig. \ref{fig5}), the decoherence
increase the damping of the oscillation as expected. And at last,
in Fig.\ref{fig6}, we have plotted the dynamics of the condensate
in the regime between the quantum tunneling and self-trapping. We
see that the decoherence increases the tunneling at the beginning
of evolution, and then destroys the quantum
tunneling/self-trapping after a few cycle of evolution.

In summary, we have studied the dynamics of Bose-Einstein condensate
in a double-well potential. The dynamics is govern by the master
equations with the condensate operator that comes from the
condensate-environment coupling. Two kinds of decoherence
characterized by $\sigma_+$ and $\sigma_x$ are considered. By
numerically solving the master equation, we show that there is a
jump-like change in the BEC population   due to the first kind of
decoherence($\sigma_+$). With the decoherence rate increasing, the
jump-like change in the population becomes unclear, resulting in the
decoherence-rate dependent self-trapping. When the second kind of
decoherence($\sigma_x$) dominated,   the decoherence first drives
the BEC from the self-trapping regime to the quantum-tunneling
regime, then it destroys the quantum tunneling  in the double-well
system. The limitation of this paper is that we have treated the
environment Markovian and have ignored the quantum fluctuation in
the condensate atoms. This may limit the application of the
formalism to  real double-well systems.

\ \ \\
This work was supported by EYTP of M.O.E,  NSF of China under Grant
No. 60578014.\\


\begin{references}
\bibitem{milburn97} G. J. Milburn, J. Corney, E. M. Wirght, and D. F.
Walls, Phys. Rev. A {\bf 55}, 4318(1997).

\bibitem{smerzi97} A. Smerzi, S. Fantoni, S. Giovanazzi, and S.R.
Shenoy, Phys. Rev. Lett {\bf 79}, 4950(1997).

\bibitem{raghavan99} S. Raghavan, A. Smerzi, and V. M. Kenkre, Phys.
Rev. A {\bf 60}, R1787(1999); S. Raghavan, A. Smerzi, S. Fantoni, S.
R. Shenoy, Phys. Rev. A {\bf 59}, 602(1999).

\bibitem{kohler02} S. Kohler and F. Sols, Phys. Rev. Lett. {\bf 89},
060403(2002).

\bibitem{salasnich00} L. Salasnich, Phys. Rev. A {\bf 61},
015601(2000).

\bibitem{wang06} G. F. Wang, L. B. Fu, and J. Liu, Phys. Rev. A {\bf
73}, 013619(2006); W. D. Li and J. Liu, Phys. Rev. A {\bf 74},
063613(2006); L. B. Fu and J. Liu, Phys. Rev. A {\bf 74}, 063614
(2006); B. B. Wang, P. M. Fu, J. Liu, and B. Wu, Phys. Rev. A {\bf
74}, 063610(2006).

\bibitem{likharev79} K. K. Likharev,  Rev. Mod. Phys {\bf 51},
101(1979).

\bibitem{pereverzev97} S. V. Pereverzev, A. Loshak, S. Backhaus, J.
C. Davis, and R. E. Packark, Nature (London) {\bf 388}, 449(1997).

\bibitem{sukhatme01} K. Sukhatme, Y. Mukharsky, T. Chui, and D.
Pearson, Nauture (London) {\bf 411}, 280(2001).

\bibitem{albiez05} M. Albiez, R. Gati, J. F\"olling, S. Hunsmann, M.
Cristiani, and M. K. Oberthaler, Phys. Rev. Lett. {\bf 95},
010402(2005).

\bibitem{dalvit00} D. A. R. Dalvit, J. Dziarmaga, and W. H. Zurek,
Phys. Rev. A {\bf 62}, 013607(2000).

\bibitem{louis01} P. J. Y. Louis, P. M. R. Brydon, and C. M. Savage,
Phys. Rev. A {\bf 64}, 053613(2001).

\bibitem{micheli03} A. Micheli, D. Jaksch, J. I. Cirac, and P.
Zoller, Phys. Rev. A {\bf 67}, 013607(2003).

\bibitem{huang06} Y. P. Huang and M. G. Moore, Phys. Rev. A {\bf
73}, 023606(2006).


\end{references}
\end{document}